\documentclass[final,5p,times,twocolumn]{elsarticle}

\usepackage{graphicx}
\usepackage{amssymb,amsmath}

\usepackage{hyperref}

\usepackage{lineno}

\usepackage{xcolor}
\newcommand{\lj}[1]{#1} 
\newcommand{\rh}[1]{#1} 
\newcommand{\al}[1]{#1} 

\biboptions{sort&compress}

\journal{Current Opinion in Colloid and Interface Science}

\newcommand{\debye}{\lambda_\text{D}}

\newcommand{\lgc}{\ell_\text{GC}}

\begin{document}

\begin{frontmatter}


\title{Measuring surface charge: why experimental characterization and molecular modeling should be coupled}


\author[dpe]{Remco Hartkamp}
\author[ilm]{Anne-Laure Biance}
\author[ilm]{Li Fu}
\author[cea]{Jean-Fran\c{c}ois Dufr\^{e}che}
\author[ilm]{Oriane Bonhomme}
\ead{oriane.bonhomme@univ-lyon1.fr}
\author[ilm]{Laurent Joly}
\ead{laurent.joly@univ-lyon1.fr}


\address[dpe]{Process \& Energy Department, Delft University of Technology, Leeghwaterstraat 39, 2628 CB Delft, The Netherlands}
\address[ilm]{Univ Lyon, Universit\'e Claude Bernard Lyon 1, CNRS, Institut Lumi\`ere Mati\`ere, F-69622 Villeurbanne, France}
\address[cea]{Institut de Chimie S\'eparative de Marcoule ICSM, UMR 5257 CEA-CNRS-ENSCM-Universit\'e Montpellier, B\^atiment 426, F-30207 Bagnols-sur-C\`eze, France}

\address{\copyright 2018. This manuscript version is made available under the CC-BY-NC-ND 4.0 license \url{http://creativecommons.org/licenses/by-nc-nd/4.0/}\\ 
DOI: \url{https://doi.org/10.1016/j.cocis.2018.08.001}}

\begin{abstract}
Surface charge controls many static and dynamic properties of soft matter and micro/nanofluidic systems, but its unambiguous measurement forms a challenge. 
Standard characterization methods typically probe an effective surface charge, which provides limited insight into the distribution and dynamics of charge across the interface, and which cannot predict consistently all surface-charge-governed properties. 
New experimental approaches provide local information on both structure and transport, but models are typically required to interpret raw data. 
Conversely, molecular dynamics simulations have helped showing the limits of standard models and developing more accurate ones, but their reliability is limited by the empirical interaction potentials they are usually based on. 
This review highlights recent developments and limitations in both experimental and computational research focusing on the liquid-solid interface. Based on recent studies, we make the case that coupling of experiments and simulations is pivotal to mitigate methodological shortcomings and address open problems pertaining to charged interfaces. 
\end{abstract}


\end{frontmatter}


\section{Introduction}
\label{S:1}

When a solid surface meets an aqueous electrolyte, physical or chemical mechanisms can generate an electric surface charge \cite{Andelman1995,IsraelachviliBook,Hunter2001,LyklemaBook}. Ions in the liquid reorganize to form a nanometric layer to balance the surface charge, the electrical double layer (EDL). Surface charge governs the stability and dynamics of soft matter systems, and as such it is a key property to characterize. Surface charge also drives the response of nanofluidic systems to thermodynamic gradients \cite{Anderson1989}. The development of new membranes to harvest e.g. blue energy (the osmotic energy of sea water) \cite{Siria2013,Feng2016,Siria2017} has led to a renewed interest for finding new functional interfaces with optimal surface charge. 

In that context, however, it is not clear that all interfacial properties governed by surface charge can be described with a single, well-defined quantity. For instance, equilibrium interactions between colloids depend on the static distribution of ions in the EDL, but their electrophoretic motion also depends on interfacial hydrodynamics \cite{Delgado2007,Lyklema2011}. Similarly, different types of osmotic flows \cite{Anderson1989} and surface conductivity \cite{Lyklema2001} might be controlled by a differently defined surface charge. Consequently, the results of standard characterization methods might not be easily used to simultaneously predict all properties governed by surface charge. 

In this short review, we will first discuss the difficulty in defining surface charge. We will then give a brief overview of standard models and experimental tools (detailed descriptions can be found in textbooks or reviews, e.g. \cite{Andelman1995,IsraelachviliBook,Hunter2001,LyklemaBook}), and discuss their limits. Next, we will give an overview of recent developments in experimental characterization, and show how progress in molecular modeling has transformed our understanding of the EDL. Finally, we will discuss benefits of coupling state-of-the-art experimental tools with molecular modeling to obtain a comprehensive picture of the interfacial structure and dynamics, necessary to accurately predict all surface-charge-related properties of liquid-solid interfaces. 
A complementary point of view can be found in a recent review focused on water at interfaces \cite{Bjorneholm2016}.

\section{Surface charge: an ill-defined concept}

\lj{The concept of surface charge cannot be defined without ambiguity because } 
liquid-solid interfaces are globally uncharged, with any surface charge being compensated by an oppositely charged EDL. 
\lj{It is therefore a question of separating charges between the surface and the liquid. However, } 
both the charged species at the surface and those in the EDL can have complex structure and dynamics. 
For instance, interfacial charged species can be strongly bound to the solid, or free to diffuse along the surface \cite{Maduar2015}. Furthermore, ions in the EDL can have a reduced mobility, or belong to a hydrodynamic stagnant layer. 

From this complex atomistic picture, different effective surface charges can be defined and measured, which quantify different physical phenomena at a larger scale. First, interactions between solid surfaces in solution result from the long-range distribution of ions between the surfaces. This can be used to define a static surface charge, which will for instance control the stability of colloidal systems. One can also define dynamic surface charges. In particular, the electroosmotic mobility can be used to define an electrokinetic charge \cite{Delgado2007,Lyklema2011}. However, it is not clear that the same electrokinetic charge can also describe other osmotic flows, e.g., diffusioosmosis and thermoosmosis. Finally, surface conductivity \cite{Lyklema2001} can be used to define yet another surface charge. 

Therefore, a suitable characterization method has to measure the relevant effective surface charge corresponding to a given phenomenon, or to provide a detailed enough description of the interface, which can be used to evaluate the adequate effective surface charge. In the second case, an accurate model of the interface is also needed. 
%
The discrepancy between surface charge measurements, and the use of macroscopic theories -- which are not \textit{a priori} justified at the nanometer scale -- to interpret them, underline the needs of coupling experimental measurements to molecular modeling in order to take proper account of the concept of surface charge.
In the following section, we will review the standard models of the EDL, the standard characterization tools based on these models, and highlight their limits.

\section{Standard approaches}

\subsection{Standard models}

\begin{figure}
\centering\includegraphics[width=0.65\linewidth]{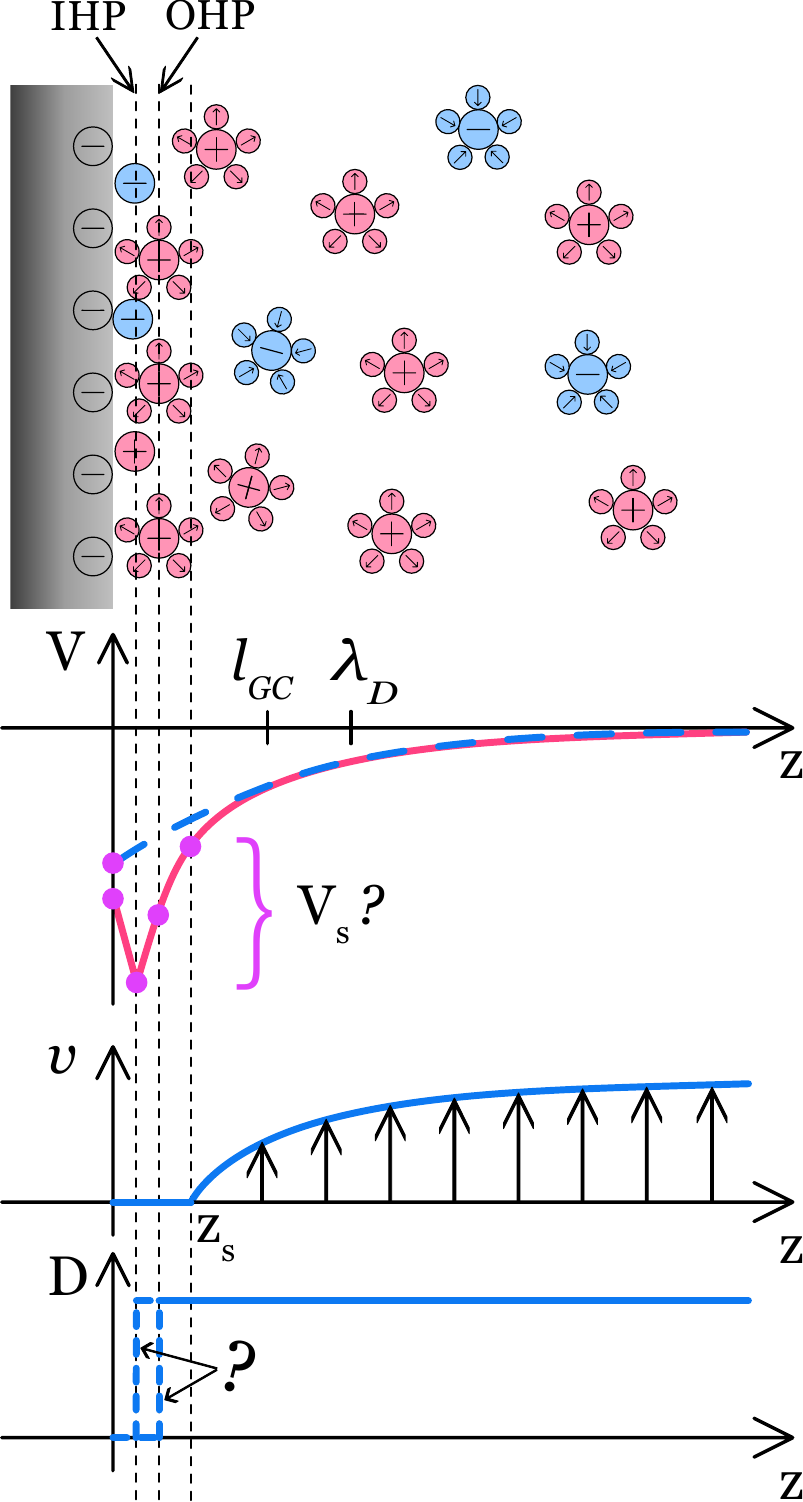}
\caption{Standard model of the electrical double layer. From top to bottom: ion distribution; electric potential profile $V$ (full red line: true potential; dashed blue line: apparent exponential potential as seen far from the interface); velocity profile $v$; ion diffusion coefficient profile $D$. Even from this traditional description, a number of surface potentials $V_\text{s}$ \lj{(represented by magenta points) } and corresponding surface charges can be defined, see text for details.} 
\label{fig:standard_model}
\end{figure}

A comprehensive model of a charged interface in an aqueous electrolyte (but even pure water is one) needs to describe the distribution of charge, the dynamics of charged species, and interfacial hydrodynamics (see Fig. \ref{fig:standard_model}). 

For the charge distribution, models generally separate the EDL into two regions. Beyond a few molecular sizes of the solid surface, ions can be considered as point charges and one commonly uses the mean-field Poisson-Boltzmann (PB) equation to describe their distribution, assuming that the dielectric permittivity of the solvent is local, isotropic and homogeneous \cite{Andelman1995}. This equation predicts that the electric potential and the local charge density decrease exponentially with the distance from the charged interface. The decay range is given by the Debye length $\debye$, which scales as the inverse square root of the salt concentration (Fig. \ref{fig:standard_model}). 
The PB equation also predicts that the charge of the EDL can concentrate in a region thinner than the Debye length with a non-exponential decay. Specifically, this happens when the so-called Gouy-Chapman length $\lgc$, which scales as the inverse of the surface charge, becomes smaller than $\debye$. The charge of the EDL then concentrates in a region of thickness $\lgc$ (Fig. \ref{fig:standard_model}). 

Very close to the surface (a few molecular sizes), \lj{the hypotheses underlying the PB equation are especially poorly justified. } 
One generally introduces the so-called Stern layer to describe this region \cite{Stern1924,Grahame1947}. 
The Stern layer is usually assumed to consist of adsorbed ions -- specifically or not, which may be partially or fully dehydrated. A number of planes and layers are defined accordingly, e.g., the inner Helmholtz plane (IHP), below which ions are specifically adsorbed and at least partially dehydrated, and the outer Helmholtz plane (OHP) separating the adsorbed hydrated ions and the diffuse layer obeying the PB equation (Fig. \ref{fig:standard_model}). It has also been recognized that the dielectric permittivity of the solvent below the IHP can deviate from that beyond the IHP, \rh{due to a preferred } orientation of solvent molecules in response to surface charge \cite{Bockris1963}. Finally, theoretical models beyond the standard PB framework have been developed to account explicitly for, e.g., correlations, image charges, finite-ion-size effects, or specific adsorption in the Stern layer \cite{Andelman1995,IsraelachviliBook}.   

For hydrodynamics, a local and homogeneous viscosity is usually assumed, together with a no-slip boundary condition. A stagnant layer is often introduced, defining the shear plane where the hydrodynamic velocity vanishes (Fig. \ref{fig:standard_model}). While the stagnant layer does not participate to the flow, whether its diffusion dynamics is bulk-like, hindered or even frozen remains unclear and may depend on the specifics of the interface. 
\lj{The hydrodynamic shear plane position $z_\text{s}$ is not simply related to the ``static'' IHP and OHP. However, $z_\text{s}$ is commonly confused with the IHP or OHP, i.e., the stagnant layer and the Stern layer are assumed to share the same boundaries.} 

Finally, regarding ion mobility, standard models assume bulk-like diffusion for ions in the diffuse layer \cite{Bikerman1932,Lyklema2001}, and immobile ions in the Stern layer, although a dynamic Stern layer is sometimes introduced to explain anomalies in surface conductivity \cite{ZukoskiIV1986I,ZukoskiIV1986II,Rosen1993,Netz2003}. 

Standard models provide an effective description of the interface, introducing a limited number of adjustable parameters with a simple physical interpretation. 
These models can relate macroscopic experimental measurements to microscopic parameters such as surface potential or shear plane position. They can also be adjusted or extended to consistently describe different effective surface charges, for instance the electrokinetic charge or the surface conductivity. However, being fundamentally effective models, which are not based on an atomistic description of the interface, they cannot be expected to provide a comprehensive and consistent prediction of all the different effective surface charges.

\subsection{Standard experimental tools}

\begin{figure}
\centering\includegraphics[width=0.9\linewidth]{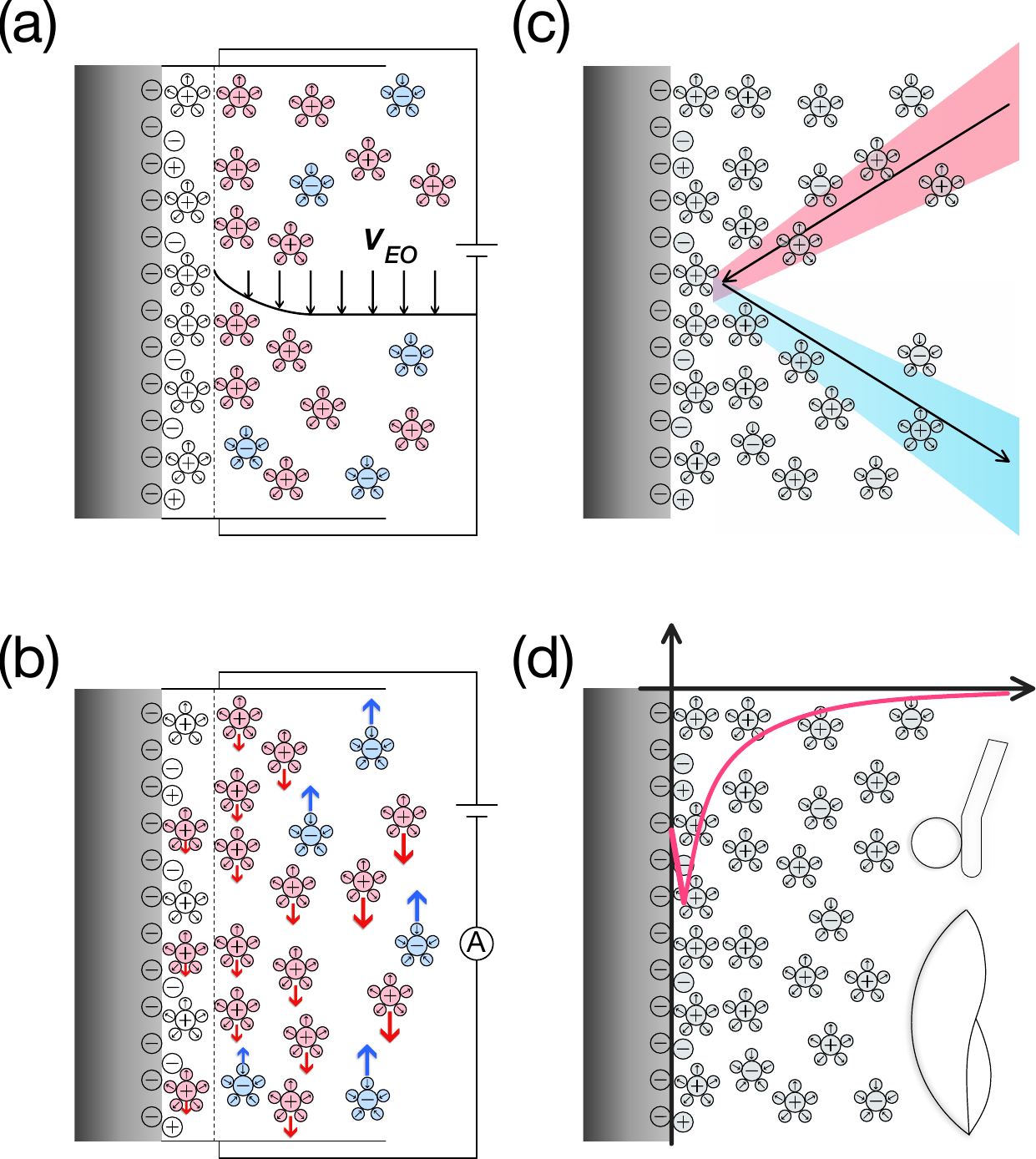}
\caption{Different experimental tools: (a) electrokinetic transport; (b) surface conductivity; (c) X-ray reflectivity or nonlinear optics; (d) AFM/SFA-like tools. Methods illustrated in (a) and (b) only provide information on ions highlighted in color, while methods illustrated in (c) and (d) are non-specific.} 
\label{fig:exp_figure}
\end{figure}

On standard basis, different methods are used to characterize surface charge. Some are based on transport properties (Fig. \ref{fig:exp_figure}a-b), and in particular electrokinetic characterization of colloids, porous materials or surfaces. Others rely on probing static properties of the interface, such as electrostatic potentials (Fig. \ref{fig:exp_figure}d). 

\subsubsection{Transport measurements}

Electrokinetic methods have been widely used to investigate the mobile part of electrolyte solutions at the interface with suspended colloids or in micro/nanochannels. In these methods, for example the electrophoretic mobility is measured, from which an electrokinetic potential, or $\zeta$-potential, is inferred \cite{Delgado2007}. 
Alternatively, an electroosmotic plug flow in micro/nanochannels can be determined, using for example the all-electric current monitoring method \cite{Huang1988} or some fluorescent probe \cite{Santiago1998}. According to Onsager reciprocal relations, the $\zeta$-potential can also be determined from the streaming current \cite{vanderheyden2005}. 
Within the standard description of the EDL, the $\zeta$-potential can be identified with the value of the electric potential at the shear plane, separating the stagnant layer from the flowing liquid \cite{Delgado2007}, see Fig. \ref{fig:standard_model}. 
This electrostatic potential can in turn be related to the electrokinetic surface charge, \textit{i.e.}, the total charge contained behind the shear plane, thanks to the Grahame equation \cite{Grahame1947}. \lj{This equation is based on the mean-field PB description of the diffuse EDL, so that  
it relies on the assumptions that the shear plane is located outside the Stern layer and that the dielectric permittivity does not vary beyond the shear plane. } 
Another method is to probe the conductivity in porous materials or nanopores at low salt concentration \cite{Hunter2001,Stein2004,Li2016surfcond}. In this regime, a so-called surface conductivity is measured, which is a signature of the electrostatic environment in the vicinity of the surfaces. \lj{However, surface conductivity also depends on ion mobility close to the surface, and can include contributions from the solid or from quantum charge transport at the interface, so that its relation to the bare surface charge is complex. } 

\lj{Generally, } electrokinetic experiments provide no direct insight into the bare surface properties of a colloid or channel wall, to which ions are adsorbed. 
Moreover, despite the good match between results obtained from streaming current or electroosmosis experiments, a discrepancy with the surface charge obtained from surface conductivity remains \cite{Rosen1993,Kijlstra1993,Rasmusson1995,Rasmusson1997a}, underlining the complexity of the systems when transport properties are considered.

\subsubsection{Static measurements}

Static measurements typically either measure the electrostatic surface potential, or the number of ionized groups at the surface, to determine the surface charge. 

Surface potential measurements have been expended in the 1990s by measuring the forces involved between a sphere and a plane using AFM/SFA-like tools \cite{Ducker1991,Atkins1993,Larson1993,Li1993,Biggs1994}, see Fig. \ref{fig:exp_figure}d. 

The static surface charge can be measured in various ways.  
For example, potentiometric titration is commonly used to probe the bare charge of a solid surface \cite{Lutzenkirchen2012}. 
Other seminal techniques involve specific adsorption of organic conjugated compounds onto the charged surface, which is titrated using UV-Vis spectroscopy \cite{Atkins1992}. 
Similarly, neutron reflectivity is a powerful tool to determine non-selectively species adsorption \cite{Lu1998} or water structure \cite{Wang2014} at interfaces. 

Attempts to probe both electrokinetic transport properties and static properties for the same sample remain scarce due to the technical constraints of the different measurements \lj{\cite{Audry2010}}. 
%
%
\lj{For colloids, the relation between static charge (obtained by titration) and electrokinetic charge depends on the nature of the surface charge \cite{Attard00}: while electrokinetic and static potentials are consistent for AgI, for which the charge results from an excess of the ionically bonded constituents, the electrokinetic potential is lower than the static potential for metal oxides, for which the charge arises from the dissociation of hydroxyl groups. }

\section{Latest innovations in experimental characterization}

Recent innovations in determining the structure of liquid-solid interfaces focused on probing the chemical nature and position of ions in the vicinity of the interface \cite{Perera2015}, see Fig. \ref{fig:exp_figure}c. Among them, specific set-ups based on X-ray reflectivity represent a powerful tool to investigate the interfacial ion distribution, for example in the case of \rh{electrolytes } in the vicinity of electrodes \cite{Steinruck2018}. 

Thermodynamic adsorption energies have been measured using resonant anomalous X-ray reflectivity \cite{Lee2013,Bellucci2015}, and specific adsorption of chloride versus iodide near hydrophilic surfaces has been observed with X-ray standing waves \cite{BenJabrallah2017}. 
A more sophisticated method based on X-ray photoelectron spectroscopy applied to a microjet containing silica nanoparticles was used to determine precisely the influence of ion specificity on the surface potential and on the Stern layer composition \cite{Brown2016a}.   
This method has also been compared to other, more standard, surface potential determination \cite{Gmur2016}. However, most of the experimental results in the above-mentioned studies require support of modeling (density functional theory \cite{Lee2013,Bellucci2015}, molecular dynamics \cite{BenJabrallah2017}, analytical \cite{Brown2016a}) to get a full picture of the processes taking place at the interface. 
\al{Moreover, X-ray based techniques are limited because they can only identify the chemical nature of the ion and not its ionization state and electrostatic environment. } \lj{For instance, these techniques are blind to hydroxyde and hydronium ions, which have a strong impact on the structure and dynamics of charged interfaces. }

The electrostatic potential of interfaces has also been probed using nonlinear optics, sum frequency generation (SFG) \cite{Jubb2012,Lovering2016,Schaefer2017} -- evidencing a cationic specific Stern layer structure -- and second harmonic generation (SHG) \cite{Lutgebaucks2016}. 
Furthermore, surface potential measurements in recent ion-sensitive field-effect transistor (ISFET) studies have provided new insight into the effect of pH and salinity of various electrolytes on the EDL structure and found that standard complexation models cannot explain the observed behavior \cite{Tarasov2012,Parizi2017,Sivakumarasamy2018}. 
Here also, molecular simulations have been used to interpret the results \cite{Fitts2005,Hosseinpour2017}. 

More direct probing techniques have also been used to gain insight into the interfacial fluid structure.  
Following huge improvement of AFM resolution, it became possible to probe the structure of the interface \cite{Siretanu2014, Zhao2015Remco,Liu2018}, and even the residence time of single ions in the Stern layer \cite{Ricci2013,Ricci2017}.

Radically different approaches also focused on the determination of ion repartition at interfaces using transport property characterization. Beyond standard electrokinetic characterization, new electrokinetic properties have been investigated, and in particular the diffusiophoretic \cite{Abecassis2008, Shi2016} and diffusioosmotic \cite{Lee2014osmotic,Siria2013} response to ion concentration gradients in solution. 
Whereas the $\zeta$-potential measured from diffusioosmotic and electroosmotic velocities coincide \al{for KI, LiI and NaI salts in the vicinity of a silica interface } \cite{Lee2014osmotic}, discrepancies are observed when considering the diffusioosmotic current \rh{of a KCl solution flowing through a membrane formed by boron nitride nanotubes } \cite{Siria2013}. In the latter, the surface potential is in good agreement with the one measured by surface conductivity, rather than the one calculated from electrokinetic mobility. A similar discrepancy between the $\zeta$-potential and the thermophoretic mobility was recently highlighted \rh{for functionalized polystyrene particles } \cite{Burelbach2017a}.  

To disentangle the couplings between static and transport properties of interfacial ions, a few attempts to measure both the electrokinetic response of the interface together with its structure have been documented \cite{Scheu2013}. For example, Jalil and Pyell \cite{Jalil2018} combined their own standard electrokinetic measurements with X-ray spectrometry data of Brown et al. \cite{Brown2016a} to get a more refined picture of the EDL \rh{for monovalent electrolyte solutions in contact with silica nanoparticles. }
In a more direct approach, in situ SFG experiments at liquid-solid interfaces under flow have shown a signature of the flow on the surface potential \cite{Lis2014}. 
\lj{Conversely, SHG experiments performed at liquid-gas interfaces evidenced no modification of the signal when electroosmotic flow was generated \cite{Blanc2018}. Recent experiments coupling SHG and streaming potential at hydrophilic and hydrophobic solid-liquid interfaces also did not observe any effect of the flow \cite{Lutzenkirchen2018}. } 
These contradicting observations underline the complexity of the mechanisms involved and indicate that the relation between the EDL structure and dynamics depends on the chemical nature of the interface.

\section{Latest developments in molecular modeling}

Whereas state-of-the-art experiments revealed new information on the structure and dynamics of the EDL, molecular simulations have highlighted the limits of standard models \cite{Knecht2008,Pagonabarraga2010,Lyklema2010,Rotenberg2013,Yoshida2014,Dewan2014,Lowe2018,Uematsu2018a}, and can help refining models using a bottom-up approach. 
Molecular dynamics (MD) simulation is a particularly powerful tool to explore the structure and dynamics of the EDL. 
MD provides an explicit description of the atomic structure of the liquid-solid interface, with its time evolution computed based on empirical interaction potentials between atoms. 
MD simulations provide accurate control over environmental conditions and full access to microscopic information that is inaccessible in experiments. 
As such, simulations can be used to explain experimental observations, or to improve models and assumptions for interpreting experimental measurements \cite{Nagata2016a}. 
Indeed, the suitability of \lj{standard models } becomes questionable when screening lengths compare with the molecular size, \rh{as well as typical values for surface roughness. } In fact, both the Debye length and the Gouy-Chapman length can easily reach one nanometer for realistic salt concentrations ($> 10^{-2}$\,M) or surface charges ($> 40$\,mC/m$^2$). In that situation, \lj{standard models } can fail to describe the EDL in many ways and a more sophisticated description is needed. 

First, it is possible that the ion distribution does not follow the mean-field Boltzmann law, especially when only electrostatic interactions are taken into account. Here, MD can provide information on specific interactions that need to be included in the Boltzmann factor 
\cite{Dufreche01,Horinek2007,Huang2007,Huang2008,Ben-Yaakov2011,Calero2011,Cazade2014,Hartkamp2015,Hocine2016,Uematsu2018}. 
Furthermore, ion-ion correlations, which are particularly important with multivalent species and concentrated solutions, can strongly affect the ion distribution, and even reverse the apparent surface charge as seen far from the interface \cite{Grosberg2002}. Also here, MD can help to refine the existing models \cite{Lorenz2007,Lorenz2008,Jardat09,Siboulet2017}. 

More importantly, the dielectric permittivity of the solvent can become inhomogeneous and anisotropic near the interface, or the local permittivity approximation can break down \cite{Bonthuis2011,Zhu2012,Bonthuis2013,Zhang2013a,Parez2014,DeLuca2016,Uematsu2018}. 
Extended theories need to be used in those cases, although it has been shown that simple effective models, with just a step in permittivity at a given distance from the wall, can reproduce \lj{static and electrokinetic charges obtained by } molecular simulations (which make no assumption on the dielectric permittivity) \cite{Bonthuis2013}. \lj{Whether such simple models can also predict the surface response to other thermodynamic gradients (e.g. osmotic or thermal) remains to be explored. } 

The standard picture for hydrodynamics, with a stagnant layer followed by a liquid with constant viscosity, has also been questioned by MD simulations at the nanoscale \cite{Bocquet2010,Hansen2015}. First, the no-slip boundary condition fails on some surfaces \cite{Bocquet2007}. 
The electrokinetic mobility is then amplified by hydrodynamic slip for a given surface charge \cite{Marry03bis,Joly2004,Dufreche05bis}. 
In turn, the amplitude of slip depends on surface charge \cite{Joly2006,Huang2008,Botan2013,Jing2015}. 
Corrected continuum descriptions \cite{Muller1986,Stone2004} describe MD results well \cite{Joly2004,Joly2006}, and the amplification effect of slip was confirmed by two independent experiments \cite{Bouzigues2008,Audry2010}. 
Hydrodynamic slip is for instance key to understanding anomalous electrokinetic charge in foam films \cite{Joly2014,BarbosaDeLima2017}. Other osmotic flows can be strongly affected by slip \cite{Ajdari2006,Morthomas2009,Fu2017}, and more generally by nanoscale dynamics \cite{Lee2017}. 

\lj{Second, the stagnant layer concept has been questioned by MD results. For instance, no stagnant layer was observed on amorphous silica \cite{Siboulet2017}, beyond a monolayer of strongly adsorbed water molecules, whose thickness was comparable to the roughness of the disordered surface.  
Zhang et al. \cite{Zhang2011} also did not observe a stagnant layer on silica, but instead they showed that viscosity increased smoothly near the surface. 
Smoothly increasing viscosity profiles were also observed by others near hydrophilic surfaces \cite{Bonthuis2013,Hartkamp2015,Predota2016}, but not near hydrophobic surfaces \cite{Bonthuis2013}. 
More generally, 
various simulation studies have indicated that viscosity can be inhomogeneous  \cite{Predota2007,Knecht2013,Hartkamp2015,Predota2016,Uematsu2017} and even nonlocal \cite{Todd2008}. 
Continuum theory taking the viscosity profile into account can predict electroosmotic flow rates \cite{Zhang2011}. The complex viscosity profiles can also be described through effective sharp boundary models, for instance by a constant viscosity combined with a few angstroms thick stagnant layer on silica \cite{Zhang2011}, by a constant viscosity on slipping surfaces, or by the inclusion of a step in viscosity on non-slipping surfaces \cite{Bonthuis2013}. 
Although such simple descriptions can be convenient, their parametrization hinges on detailed insight into the interfacial region. Moreover, because these effective descriptions do not correspond to the real microscopic picture, their transferability to describe all surface-charge-governed properties is not guaranteed. } 

MD simulations have also given new insight into the diffusion dynamics of ions in the different subsections of the EDL \cite{Bouhadja2018}. 
In their pioneering work, Lyklema et al. \cite{Lyklema1998} used MD to confirm the emerging picture of a stagnant layer behaving like a gel, in which the ions can diffuse almost freely, but which does not flow globally.  
This picture explains the large possible differences between the surface charge one can extract from electrokinetic or surface conductivity measurements \cite{Rosen1993,Kijlstra1993,Rasmusson1995,Rasmusson1997a}. 
Recent MD simulations on amorphous silica \cite{Siboulet2017} found that the standard decomposition of the EDL into a Stern and a diffuse layer was inadequate. The EDL was instead decomposed into a mobile and an immobile ion population, of which the distribution overlaps. The diffusion coefficient of free ions continuously decreases close to the wall, an effect that can be described by continuum hydrodynamics \cite{HappelBrennerBook}, and can dramatically affect surface conductivity. 
Finally, on hydrophobic surfaces where the surface charge is carried by specifically adsorbed ions, the surface mobility of these ions can affect the electrokinetic response of the interface \cite{Maduar2015}.  

Even though simulations have proven valuable in the study of EDL properties, classical MD simulations are reaching their limits, because of two strong weaknesses. First, interactions between atoms are based on empirical force fields. By definition, these force fields are built to reproduce a given set of data, and their transferability to different situations is questionable. Specifically, most standard force fields are designed to reproduce equilibrium structural properties of bulk systems, and there is no guarantee that they can correctly describe the dynamical and transport properties of interfaces. 
A striking example concerns the effect that ions have on water diffusion and viscosity \cite{Marcus2009}. While some salts enhance water diffusion, most empirical force fields predict a systematic decrease in diffusion with increasing ion concentration \cite{Kim2012,Ding2014}. However, recently developed force fields have been able to qualitatively reproduce the effect of large ions \cite{Kann2014}, and to quantitatively describe the effect of small ions \cite{Yao2014,Yao2015,Li2015a} on water diffusion. 
 
The second major weakness of force field-based simulations is related to the systems described. Indeed, the atomic wall structure and charge distribution need to be constructed before the simulation can be run, often from limited information. 
Notably, the surface charge is usually imposed by assigning partial charges to the atoms in the substrate, conform the force field employed. These partial charges are kept constant throughout a classical simulation -- rendering the bare surface charge unaffected by the rearrangement of interfacial ions or solvent molecules.
Alternatively, surface reactivity is considered by using reactive force fields \cite{doi:10.1021/jp004368u,Raymand2011,Kim2013,Senftle2016} and ab initio methods \cite{Gillan2016,Nagata2016,Chen2017}. 
Such reactive simulations are important to investigate, for example, charge regulation \cite{Markovich2016,Trefalt2016}, which occurs in narrow channels due to overlapping EDLs. 

Ab initio molecular dynamics (AIMD) has been used to explore the structure and dynamics of water-oxide interfaces \cite{Masini2002,Criscenti2006,Tielens2008,Kumar2009,Liu2010,Cimas2014,Gaigeot2012,Sulpizi2012,Tazi2012a,Wesolowski2012,Tocci2014,Laporte2015,Pfeiffer-Laplaud2016,Pfeiffer-Laplaud2016a,DelloStritto2016,Reocreux2018}. In particular, AIMD can be used to compute vibrational spectra and non-linear optical response \cite{Kumar2009,Zhang2011a,Wan2013,Thomas2013,Sulpizi2013,Khatib2016}, and as such is a key tool to help interpret experimental observations.  
Recently, liquid-solid friction has been characterized with AIMD \cite{Tocci2014a,Joly2016}, opening perspectives for the investigation of other transport properties with these methods. 
At present, however, ab initio methods are typically based on density functional theory, and take electronic exchange and correlations into account through an approximate functional. This limits the accuracy of ab initio methods to describe the structure and dynamics of water-solid interfaces \cite{Gillan2016}. Most AIMD works also do not take quantum nuclear effects into account. Finally, 
the computational cost of reactive and ab initio simulations is very large
-- effectively limiting the accessible simulation time and the number of atoms. 
These restrictions are currently prohibitive for studying rare events, dilute electrolytes, or for simulating a sufficiently large system to accurately represent the structure and heterogeneous charge distribution of amorphous surfaces.

Size restrictions can be mitigated drastically with primitive-model simulations, in which the solvent is included implicitly \cite{Heyes1982}. 
For instance, such an approach was used to investigate the origin of surface charge on graphene and boron nitride \cite{Grosjean2016}. 
However, Lee et al. \cite{doi:10.1021/ct4002043} found that various physical quantities that depend on the orientation of solvent molecules were not accurately predicted by the primitive model approach. Additionally, Vangara et al. \cite{Vangara2017} recently showed, by comparing explicit and implicit solvent DFT models, that both the solvent and the ions contribute to the chemical balance between surface groups and the solution.

Apart from the importance of solvent molecules and dissolved salts, various experiments have revealed intricate effects of pH on for example specific ion adsorption \cite{Darlington2017,DeWalt-Kerian2017}. Furthermore, local enhancement of proton mobility affects the Stern conductivity \cite{Chinen2012}, but may also have important consequences for surface reactivity. 
Simulations are, in principle, suitable for elucidating the molecular-level mechanisms responsible for such pH dependence, but current computing power does not permit explicitly accounting for near-neutral pH in molecular simulation because of the immense system size required. 
\lj{Interestingly, this numerical difficulty echoes the limits of experimental methods based on X-rays, which are blind to hydronium and hydroxide ions and hence unable to provide insight on the role of pH. } 

Explicit-pH simulations and large-scale quantum-based simulations are currently beyond the feasible. However, with the continuous improvement of computing power, ab initio simulations should progressively be able to tackle an increasingly large panel of problems. Meanwhile, the information obtained from small ab initio simulations is already transferred to classical simulations using ab initio-based force fields \cite{Predota2004a,Masia2008,Machesky2008,Tazi2012,Butenuth2012}.

\section{Why coupling experiments and modeling is needed, and recent attempts}

Despite recent developments in experimental characterization and molecular modeling, the study of local structure and dynamics at the solid-electrolyte interface remains restricted by the inherent limitations of the respective methods. The ambiguity in what quantity is measured by each method is a considerable limitation, making it difficult to form a unified understanding of the EDL by combining data from multiple experimental techniques. 
Ambiguity or uncertainty of measurements can even propagate when combining techniques.
For example, the charge held within the Stern layer can be estimated by combining titration experiments to determine the bare surface charge density and electrokinetic experiments to infer the charge contained in the mobile region \cite{Lyklema2011}. However, the latter relies on the assumption that the electrokinetic charge equals the charge held in the diffuse layer, i.e., 
the shear plane is assumed to coincide with the OHP.
Similarly, the difference between the static and electrokinetic potentials has been used to estimate the Stern layer thickness, assuming a constant permittivity across the Stern layer \cite{Brown2016b}. The calculated thickness depends linearly on the assumed Stern-layer permittivity and on the approximated potential drop between the surface potential and the electrokinetic potential, which were obtained using two different techniques \rh{and material samples}. Furthermore, this approach again assumes that the shear plane marks the outside of the Stern layer. 
Evidently, inferred quantities, such as the Stern layer thickness and its charge, can be highly sensitive to assumptions and models. 

On the other hand, 
simulations can help to interpret and complement experiments, without the need for assumptions or theoretical models. However, direct and quantitative comparison between experiments and simulations is often difficult for two major reasons: first, due to empirical force fields in classical MD, or approximations in ab initio methods, and second, because atomistically-detailed computations are limited to short simulation times and small systems, whereas many experiments are limited by their spatial and temporal resolution.  Direct coupling between simulations and experiments thus presents a major challenge. Yet, combining these disciplines can help to leverage their complementary strengths. Specifically, experiments are essential to validate simulation results and to improve simulation force fields, while accurate simulations are helpful to interpret and explain experimental measurements.  

Thus far, few studies have combined experiments and simulations to gain deep understanding of interfacial fluid properties \cite{Predota2004,Fitts2005,doi:10.1021/jp057096+,Labbez2007,Labbez2011,Calero2011,Nagata2012,Ricci2013,Semenov2013,Predota2013,Siretanu2014,Page2014,Wang2014,Brkljaca2015,Bellucci2015,Khatib2016,Hussain2016,Bourg2017,Perrine2017,Hosseinpour2017,Steinruck2018,Sivakumarasamy2018,Ge2017,Lutzenkirchen2018}.
For example, P{\v{r}}edota et al. \cite{Predota2013} performed MD simulations and titration experiments for different electrolytes near a hydroxylated (110) rutile surface. The simulations suggested that different slopes in the titration curves were caused by different adsorption mechanisms. 
In a recent study, Bourg et al. combined X-ray reflectivity, MD simulation, and complexation theory to provide detailed insight into the interfacial structure of 0.1\,M alkali chloride solutions on a muscovite mica surface \cite{Bourg2017}. \al{The authors showed that the structure of the first molecular layer of water was determined predominantly by interaction with the surface, whereas the structure of the second layer depended also on interactions with adsorbed interfacial ions. Water beyond the first two monolayers exhibited density layering, but showed no evidence of sensitivity to short-range interactions with either the surface or adsorbed ions. } With the exception of $\mathrm{Li}^+$, the 
experimentally and computationally measured ion exchange energies were in close agreement. The trend in the exchange energy of different ions suggested that not only the hydration free energy was important, but also the match between the surface structure and the hydration structure of the ions. 

Combined experiments and simulations were also used to understand the mechanisms underlying charge inversion \cite{Grosberg2002}. 
Labbez and coworkers combined surface titration measurements, electrophoretic experiments, and implicit-solvent grand canonical Monte Carlo simulations to study the charging behavior of calcium silicate hydrate \cite{doi:10.1021/jp057096+,Labbez2011}. 
The authors found that the apparent charge inversion observed for concentrated divalent solutions decreased, or even disappeared, upon the addition of monovalent electrolytes to the electrolyte mixture. This contrasts the idea that charge inversion increases with ion concentration. Using MD simulations and electrophoresis experiments, Calero et al. observed charge inversion for large organic monovalent ions near a hydrophobic colloid, but not near a hydrophilic surface \cite{Calero2011}. This demonstrated that charge inversion can occur also when ion-ion correlations are negligible. 
Semenov et al. \cite{Semenov2013} calculated the electrokinetic mobility of a single latex colloid in a trivalent electrolyte solution from implicit-solvent MD simulations combined with hydrodynamic theory. Electrostatic and specific adsorption were both essential to predict the mobility reversal observed in their optical tweezer experiments. 

The insights obtained in the studies described above could not be obtained solely using experiments, due to the need for non-invasive measurements with a sub-nanometer resolution, or detailed insight into the interactions between individual atoms. On the other hand, only modeling the problem would also be unsatisfactory, as the validity of the results is not guaranteed.

\section{Conclusions and outlook}

A complete understanding of the dynamics and structure of ions repartition near surfaces requires first a full experimental investigation. 
Most of the techniques nowadays rely on determining charge transport properties (probing electrokinetics mainly) \emph{or} static properties such as surface potential or ion repartition structure. However, most recent advances show that both are intimately coupled; transport can indeed change the ion repartition and \textit{vice versa}. Only a full characterization of the liquid structure near the interface while electrokinetic transport takes place, under a range of conditions, would allow giving realistic inputs in this subject. Another direction that needs to be tackled is the identification of specific solid systems. Indeed, most oxide surfaces, especially silica, are complex: their properties depend a lot on the preparation and despite huge efforts in developing methodology in the experimental community, results are poorly reproducible from one group to the other. Identifying robust model materials would be a real asset.

Molecular simulations can be a valuable tool in the quest for identifying suitable model materials and the coupling between static and transport properties. 
Leveraging the strengths of AIMD, and combining this with computationally cheaper tools, may hold much promise for the future of computational molecular research. In fact, AIMD has adopted an increasingly important role in the study of molecular fluids and fluid-solid interfaces in recent years. 
The main drawback of this technique has thus far been its large computational cost, which strongly limits the accessible simulation time and system sizes. 
Although the accessible time scale in AIMD is insufficient to directly probe transport properties, AIMD could instead be used to explore free energy profiles \cite{Baer2011,Yao2017}. 
Alternatively, the limitation of accessible time scales can be mitigated using machine learning, e.g., to predict infrared spectra \cite{Gastegger2017}. Machine learning was also key
to mitigate high computational costs  to calculate free energy differences in a classical MD system \cite{Riniker2017}. 
Machine learning, combined either with classical or quantum simulation, can be a powerful tool in the development of more versatile and transferable simulation force fields, which require optimization against a large set of conditions and variables. Particularly, fingerprint algorithms, an aspect of machine learning, were recently suggested as a `\textit{useful building block for constructing data-driven next generation force fields}' \cite{Tang2018}. 
In addition to the potential speedup to be achieved with machine learning, the arrival of quantum computing may hold promise for drastically accelerating molecular simulations. 
Finally, advanced techniques capable of taking quantum nuclear effects into account are increasingly accessible \cite{Nagata2012,Cisneros2016,Markland2017,Litman2018} and could help improving the description of water-based systems.

In conclusion, both experimental characterization and numerical modeling of charged interfaces have vastly progressed over the past years, and current developments give hope for a bright future. MD does not only help to understand the microscopic phenomena; it also helps to extend the validity of the experimental measurements. Indeed, numerous macroscopic parameters that are necessary to interpret the measurements (adsorption constant, slip length, \textit{etc.}) can be calculated from MD, so that the treatment of experimental data is facilitated. Furthermore, very often, simulations and experiments can be directly compared without requiring the use of ill-defined concepts such as $\zeta$-potential or effective surface charges. 
This does not mean that such macroscopic parameters are not useful anymore. Any macroscopic theory needs such average quantities. It means that thanks to this direct comparison between experiments and simulations, one can decide which effective quantity is relevant for a given system. 
%
\lj{New, more efficient models have been developed in recent years, but there is still room for improvement. For instance, future models could go beyond traditional assumptions of distinct layers separated by sharp interfaces. In particular, the complex 3D structure and dynamics of the EDL due to surface roughness and chemical heterogeneities could be taken explicitly into account. Such 3D models could indeed provide more insight on the discrepancy between static and dynamic surface charge, and identify parameters controlling this discrepancy. 
Coupling experiments and atomistic modeling is the best method to assess the value of recent and future models. } 

We hope we have convinced the reader that major advances in the understanding and detailed characterization of surface charge(s) will come from the coupling between state-of-the-art experiments and molecular simulations.

\section*{Acknowledgement}

This work is supported by the ANR, project ANR-16-CE06-0004-01 NECtAR. LJ is supported by the Institut Universitaire de France.

\section*{Recommended reading}

$\bullet$ of special interest

$\bullet\bullet$ of outstanding interest

\cite{Delgado2007} $\bullet$ Review on electrokinetic transport measurements, discussing the models used for their interpretation.  

\cite{Lyklema2011} $\bullet\bullet$ This article presents a critical discussion of the difference between bare surface properties and electrokinetic quantities among other things. 

\cite{Bjorneholm2016} $\bullet$ Review on latest advances in understanding the structure and dynamics of water at various interfaces. 

\cite{vanderheyden2005} $\bullet\bullet$ This work compared the streaming current using various channels and ion concentrations. 
 
\cite{Audry2010} $\bullet$ Comparison between AFM and electrokinetic measurements, showing the amplification of electrokinetic mobility by liquid-solid slip. 

\cite{Bellucci2015} $\bullet$ X-ray reflectivity and ab initio simulations are used to explore cation adsorption at the water-quartz interface. 

\cite{BenJabrallah2017} $\bullet$ X-ray standing waves are used to explore ion distribution at the water-silica interface. 

\cite{Brown2016a} $\bullet\bullet$ This work introduces X-ray photoelectron spectroscopy from a liquid microjet, which provides insight into the potential drop across the Stern layer. 
 
\cite{Siretanu2014} $\bullet$ The authors combine AFM experiments with DFT simulations to measure and explain how different electrolytes and concentrations affect the effective surface charge. 

\cite{Ricci2013} $\bullet$ This study succeeded in probing adsorbed ions with high-resolution AFM experiments, supported by MD simulation. 
 
\cite{Lis2014} $\bullet\bullet$ SFG experiments showed that surface charge can be reversibly affected by a flow, suggesting a strong coupling between chemistry and transport. 


\cite{Huang2007} $\bullet\bullet$ MD work showing anomalous electroosmosis due to ionic specificity, resulting for instance in a nonzero zeta potential for a neutral surface. 

\cite{Calero2011} $\bullet$ This study combines MD simulation and electrophoresis experiments to demonstrate that solvation free energy can cause or prevent charge inversion, depending on the colloid surface hydrophobicity. 
 
\cite{Siboulet2017} $\bullet$ This computational study reveals interesting insights by deviating from the traditional way of looking at the electrical double layer. 
 
\cite{Bonthuis2013} $\bullet\bullet$ MD work showing that both viscosity and dielectric permittivity are affected by solvent structure close to interfaces. 

\cite{Bocquet2010} $\bullet$ Very complete review on both experimental and theoretical aspects on coupled electrokinetic transport near solid interfaces. 

\cite{Joly2004} $\bullet$ MD work showing that the zeta potential depends critically on hydrodynamic boundary condition, and can be amplified by liquid-solid slip. 

\cite{Predota2016} $\bullet\bullet$ This study uses MD simulation to provide deep understanding of the origin of the macroscopically measured zeta potential. 
 
\cite{Knecht2013} $\bullet$ MD work discussing the limits of traditional models of electoosmosis in the context of lipid membranes. 


\cite{Ding2014} $\bullet\bullet$ An article illustrating the power of ab initio methods, which can describe the effects of salts on water diffusion, in contrast with force field-based simulations. 

\cite{Gillan2016} $\bullet$ A review discussing the interest and shortcomings of ab initio methods to describe aqueous systems. 


\cite{Khatib2016} $\bullet$  A recent work showing how coupling SFG and ab initio simulations can help to understand water-mineral interfaces at the molecular level. 

\cite{Tocci2014a}  $\bullet$ AIMD was used to show that liquid-solid friction can depend crucially on specific electronic structure effects. 

\cite{Grosjean2016} $\bullet$  DFT calculations were used to explore the origin of surface charge on graphene and boron nitride.

\cite{Predota2004}  $\bullet$ A combined MD and X-ray study of ion adsorption at the water-rutile interface

\cite{doi:10.1021/jp057096+} $\bullet$ This study combined  simulations with potentiometric titration and electrophoresis experiments, as well as theory, to study the effect of pH and ion concentration of surface charging. 

\cite{Nagata2012}  $\bullet$ A combination of SFG spectroscopy and path integral MD simulations showed that the molecular structure of water-vapor interfaces is affected by nuclear quantum effects. 

\cite{Bourg2017} $\bullet\bullet$ The consistent views provided by the experiments and simulations in this study contribute to a better understanding of the Stern layer.

\section*{References}



\bibliographystyle{model1-num-names}
\bibliography{Mendeley.bib,refRemco.bib,al.bib,jfd.bib,refine.bib}







\end{document}